\documentclass[preprint,12pt]{revtex4-1} 

\usepackage[utf8]{inputenc}
\usepackage{graphicx}
\usepackage{amssymb}
\usepackage{amsmath}
\usepackage{lineno} 
\usepackage{url}

\newcommand{\degree}{\ensuremath{^\circ}}

\begin{document}

\title{Influence of nanoscale temperature rises on photoacoustic generation: discrimination between optical absorbers based on nonlinear photoacoustics at high frequency.}

\author{Olivier Simandoux}
\author{Amaury Prost}
\author{Jérôme Gâteau}
\author{Emmanuel Bossy}

\address{Institut Langevin, ESPCI ParisTech, CNRS UMR7587, INSERM U979, Université Paris Diderot - Paris 7, 1 rue Jussieu, 75005 Paris, France}

\begin{abstract}
In the thermoelastic regime, photoacoustic sensing of optical absorption relies on conversion from light to acoustic energy via the coefficient of thermal expansion.   In this work, we confront confront experimental measurements to theoretical predictions of nonlinear photoacoustic generation based on the dynamic variation of the coefficient of thermal expansion during the optical excitation of absorbers in aqueous solution. The photoacoustic generation from solutions of organic dye and gold nanospheres (with same optical densities), illuminated with 532 nm nanosecond pulses, was detected using a high frequency ultrasound transducer (center  frequency 20 MHz). Photoacoustic emission was observed with gold nanospheres at low fluence (a few mJ/cm$^2$) for an equilibrium temperature around $4\degree C$, where the linear photoacoustic effect in water vanishes, highlighting the nonlinear emission from the solution of nanospheres. Under the same condition, no emission  was observed with the absorbing organic dye. At a fixed fluence of 5 mJ/cm$^2$, the photoacoustic amplitude was studied as a function of the equilibrium temperature from $2\degree C$ to $20\degree C$. While the photoacoustic amplitude from the organic dye followed the coefficient of thermal expansion in water at the equilibrium temperature and vanished around $4\degree C$, the photoacoustic amplitude from the gold nanospheres remained significant over the whole temperature range. These experimental results are shown to be in qualitative agreement with theoretical calculations based on the nanoscale and nanosecond transient changes of the coefficient of thermal expansion of water around the gold nanospheres. Our results suggest that in the context of high frequency photoacoustic imaging, nanoparticles may be discriminated from molecular absorbers based on either the nonlinear fluence-dependence or the equilibrium temperature-dependence of the photoacoustic amplitude.
\end{abstract}

\maketitle

\section{Introduction}
\label{sec:Introduction}

Photoacoustics has demonstrated optical contrast imaging in biological tissues at depth beyond 1mm~\citep{Beard2011}. This non-invasive hybrid modality uses the conversion of transient illumination to ultrasound wave though thermoelastic expansion to detect optical absorption. Besides the detection of endogenous photo-absorbers such as hemoglobin and melanin, the use of exogenous contrast agents has shown to provide functional and molecular information~\citep{Ntziachristos2010}. Optical reporter agents in photoacoustics range from molecular agents to nanoparticles. Organic dyes like ICG~\citep{Buehler2010,Kim2010} and methylene blue~\citep{Song2009} are of molecular agents and have already been used to enhance visualization of the circulatory system and its dynamic. Gold nanoparticles have been used in particular to image the enhanced permeability and retention effect in tumors~\citep{Li2009} or their long term biodistribution in small animals~\citep{Su2012}. Although relative variations of the amplitude of the photoacoustic signal could be sufficient if images ar taken at different time points before and after the injection, multispectral approaches are usually employed to improve the specificity of the detection. The retrieval of spectral signature of contrast agents can be challenging. Indeed, most of reconstruction algorithms~\citep{Xu2005,Dean-Ben2012} and analytical derivation of the photoacoustic signal~\citep{Diebold1991} assume that the reconstructed initial pressure in the medium is directly proportional to the optical absorption and the fluence. However, because of the wavelength dependent absorption of the excitation before reaching the region of interest, the relationship between the fluence on the sample surface and at the imaged position can be non-trivial. Specific detection by suppressing background signal has also been demonstrated using magnetic contrast agents for magnetomotive photoacoustic imaging~\citep{Xia2012}. 

In this work, we consider photoacoustic nonlinearity as a candidate mechanism to discriminate between various type of optical absorbers. Several phenomena may induce non-linear relationships between the photoacoustic signal amplitude and the energy of the incident light, such as optical saturation~\citep{Danielli2010,Zharov2011} or temperature-dependence of thermodynamic parameters~\citep{Burmistrova1979,Calasso2001,Inkov2001}. Nonlinear phenomena could be on their own a mean of selectively detect contrast agents, similarly as what is done in the field of ultrasound imaging. In the context of biomedical imaging, one recent experimental study reported a nonlinear signal increase with the laser fluence, presumably caused by thermal coupling within aggregated nanoparticles in cells~\citep{Nam2012}. In the present study, we compare experimental results to theoretical predictions regarding the occurence of thermal-based nonlinear photoacoustic phenomena with two different types of absorbers in aqueous solution, an organic dye and gold nanospheres. The main objective of this study is to demonstrate the ability to discriminate between different types of absorbers based on thermal nonlinearity. We first give an introduction to the so called thermal nonlinearity in photoacoustics, and discuss some predictions from numerical simulations. Experimental results from \textit{in vitro} experiments are then qualitatively compared to theoretical predictions.

\section{Theoretical Background}
\label{sec:Theory}

\subsection{Nonlinear photoacoustic generation in the thermoelastic regime}
\label{subsec:PhysicalBasis}

\begin{figure}[!h]
\includegraphics[scale=1]{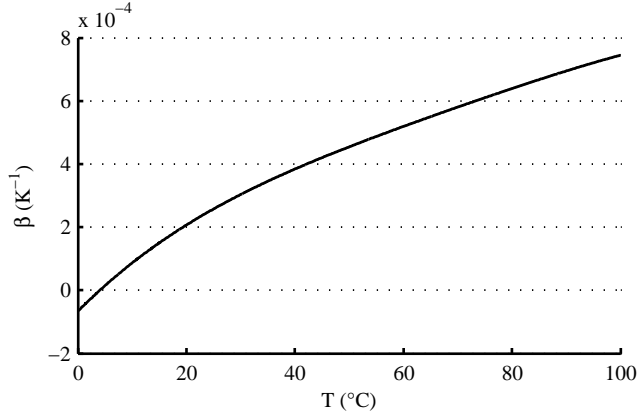}
\caption{Coefficient of thermal expansion of water $\beta$ as a function of temperature~\citep{CRC2007}. Note that $\beta$ vanishes for $T\sim4 \degree$ C.}
\label{fig:fig1}
\end{figure}

In a theoretical work published in 2001~\citep{Calasso2001}, Calasso \textit{et al}. analytically calculated a nonlinear contribution to the photoacoustic signal emitted by a point absorber. In this simplified model, the absorber immersed in a surrounding fluid was of vanishingly small size but with a finite optical absorption cross-section. As a consequence, the photoacoustic generation was dictated by the thermodynamics properties (speed of sound $c_s$, density $\rho$, specific heat capacity $c_p$ and coefficient of thermal expansion $\beta$) of the mere fluid, in particular via its Grüneisen coefficent $\Gamma=\frac{\beta c_s^2}{c_p}$. Amongst the relevant thermodynamics properties, $\beta$ shows the most significant temperature dependency, illustrated on Fig.\ref{fig:fig1}. The nonlinearity predicted in~\citep{Calasso2001} arised from the temperature-dependence of $\beta$ in the photoacoustic wave equation:
\begin{equation}
\left [\frac{1}{c_s^2}\frac{\partial^2 }{\partial t^2}-\Delta \right ] p(\mathbf{r},t)= 
\rho \beta(T) \frac{\partial^2 T}{\partial t^2}(\mathbf{r},t)
\end{equation}
where $T(\mathbf{r},t)$ is the temperature field in water. By linearizing the temperature-dependence of $\beta(T)$ around the equilibrium temperature in the medium, according to the following equation
\begin{equation}
\beta(T)=\beta(T_{eq}+\delta T)=\beta_{eq} +\delta T \frac{d\beta}{dT}(T_{eq})
\end{equation}
Calasso \textit{et al}. separately calculated two terms in the expression of the photoacoustic signal: a linear term corresponding to the photoacoustic emission assuming that $\beta$ had not changed in response to the temperature rise induced by the optical absorption, and a nonlinear term attributed to the local temperature rise. This so called thermal nonlinearity, earlier discussed in~\citep{Burmistrova1979}, results from the dynamic transient change of the coefficient of thermal expansion during the pulsed illumination. The nonlinear contribution may become significant only if temperature rises are high enough to affect the value of $\beta(T)$ during the illumination. For water, the linear contribution vanishes around $T_{eq}\sim 4\degree C$, and therefore only the nonlinear contribution should remain, as experimentally observed from protons absorption experiments in water~\citep{Hunter1981}.

In summary, thermal-based photoacoustic nonlinearity is expected to be significant in comparison to the linear contribution, and then accessible to experiments, either when the illumination fluence is large enough to induce a significant temperature rise, or when the equilibrium temperature approaches  $4\degree C$ so that the linear contribution becomes negligible as compared to the nonlinear one. The objective of the current work is to confront these qualitative predictions to experimental results performed with two types of absorbers, a solution of molecular organic dye and a solution of gold nanospheres. The temperature rise around gold nanospheres in aqueous solution may become significant, due to the large absorption cross-section of gold nanoparticles and the partial heat confinement caused by the nanometric size. On the other hand, negligible local temperature rise will be encountered with a dye solution of equivalent optical density: in this case, light is absorbed by a much larger number of dye molecules with as much lower absorption cross-sections, resulting in a very weak temperature rise at the scale of individual absorbers. Therefore, one expects thermal nonlinearity to possibly manifest itself with solution of gold nanoparticles, while dye solutions of equivalent optical density are expected to behave linearly. The following section discusses orders of magnitude involved with nanosecond illumination and heating of gold nanospheres.

\subsection{Prediction for gold nanospheres}
\label{subsec:GenerationGNS}

\begin{figure*}[!ht]
\includegraphics{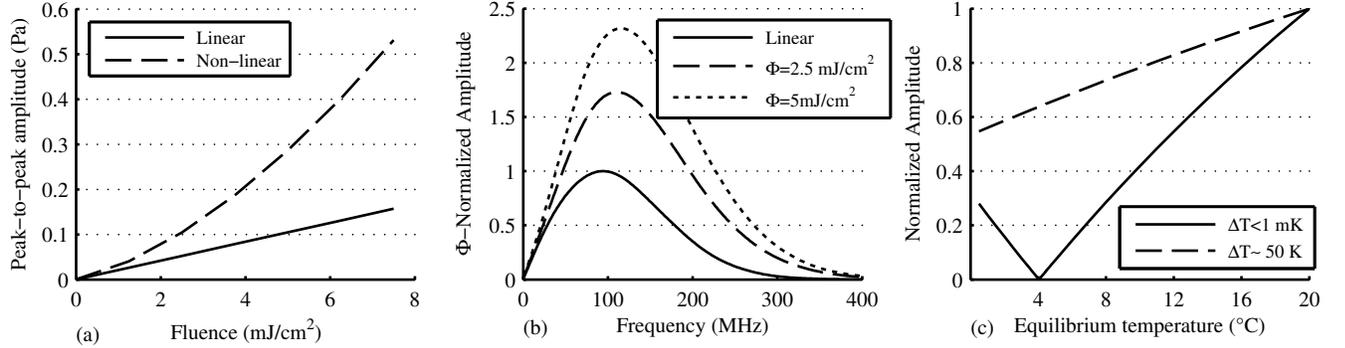}
\caption{Theoretical predictions from numerical simulations~\citep{Prost2013} on the photoacoustic emission by a single gold nanosphere (40 nm in diameter) immersed in water, illuminated by 4-ns laser pulses. (a) Peak-to-peak amplitude vs incident fluence at $T_{eq}=20\degree C$. (b) Normalized frequency spectra for three values of the incident fluence, obtained by Fourier transformation of the photoacoustic signals. Each spectra was first individually normalized by the incident fluence, before all spectra were normalized to common arbitrary units. The spectrum in continuous line corresponds to that obtained in the linear regime. (c) Evolution of the normalized peak-to-peak amplitude as a function of the equilibrium temperature, obtained in the linear (unsignificant temperature rise) and nonlinear (significant rise of the absorber temperature) regimes.}
\label{fig:fig2}
\end{figure*}

Gold nanospheres with a diameter of a few tens of nanometers cannot be reduced to point absorbers~\citep{Inkov2001}. On the other hand, it has recently been demonstrated experimentally that for nanosecond illumination, it is mostly the fluid surrounding the nanoparticles that emits photoacoustic waves~\citep{Chen2012}. From a theoretical point of view, the prediction of the photoacoustic wave emitted by a gold nanosphere requires to first solve the diffusion equation for the temperature field in the sphere and its environment, and then to use this temperature field as a source term in the photoacoustic wave equation. Because the heat diffusion caracteristic times for gold nanospheres are of the same order of magnitude as the illumination time, the problem is analytically intractable, even in the linear regime where $\beta$ is assumed to be constant. Moreover, taking into account the temperature-dependence of $\beta(T)$ makes the problem even more difficult to solve by means of analytical methods. Calasso \textit{et al}. could provide a solution only under the assumption of a point absorber and after linearization of the temperature dependence of $\beta(T)$~\citep{Calasso2001}. Taking into account the size of the absorber has been done in~\citep{Inkov2001}, but results could only be found for either thermally small or large absorbers, assumptions that do not hold for gold nanospheres. Numerical approaches are required to obtain accurate predictions of temperature rise and subsequent thermal-based photoacoustic nonlinearity. It is out of the scope in this experimental report to describe the numerical methods that can be used to solve both the thermal and the photoacoustic problems. Suffice it to say that theoretical predictions from a numerical resolution, detailed elsewhere in a publication by our group~\citep{Prost2013}, are in qualitative agreement with the main prediction that can be drawn from the work by Calasso \textit{et al}, namely that significant temperature rise around efficient optical absorbers yields a nonlinear relationship between the amplitude of photoacoustic signals and the light fluence. 

Figure~\ref{fig:fig2} summarizes additional quantitative results from~\citep{Prost2013}, which justify the experimental approach presented in this work. This figure shows results obtained for a 40-nm diameter gold nanosphere immersed in water and illuminated with a 4 ns (FWHM) laser pulse.  First, Fig.\ref{fig:fig2}(a) predicts that a relatively low fluence (typically a few mJ/cm$^2$) are needed to obtain a significant nonlinear dependence of the photoacoustic peak-to-peak amplitude with the incident fluence. Second, Fig.\ref{fig:fig2}(b) shows that the nonlinearity is predicted mostly around the centrer frequency corresponding to that of the temporal derivative of the laser pulse, i.e. approximately 100 MHz here. This feature has apparently never been mentioned nor received any attention to the best of our knowledge. It is however very important for experimental implementation as it suggests that a high frequency ultrasound detection should be used to measure thermal nonlinear effects in photoacoustics. Third, when one considers the evolution of the photoacoustic amplitude as a function of the equilibrium temperature, two different situations may be predicted, as shown in Fig.\ref{fig:fig2}(c): in the linear regime (small temperature rises, $\beta(T)\sim\beta(T_{eq})$), the temperature dependence of the photoacoustic amplitude simply reflects the temperature dependence of the coefficient of thermal expansion, as is  well known \citep{Larina2005, Shah2008}. However, when large transient temperature rises are involved ($50\degree C$ for instance in Fig.\ref{fig:fig2}(c)), i.e. when nonlinear contributions are significant, the temperature-dependence of the photoacoustic amplitude now reflects some effective value of the coefficient of thermal expansion. In particular, a significant photoacoustic amplitude is predicted around $T_{eq}\sim 4 \degree C$.  

\section{Material and Methods}
\label{sec:MM}

\begin{figure}[!h]
\includegraphics[width=0.5\textwidth]{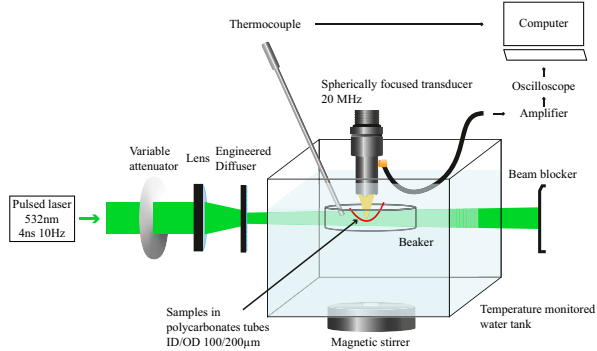}
\caption{Experimental setup}
\label{fig:fig3}
\end{figure}

An acoustic-resolution photoacoustic microscope was used to detect signals from a tube filled with absorbing aqueous solutions. The experimental setup is illustrated in Fig.\ref{fig:fig3}. Nanosecond pulses from a Q-switched laser (Brilliant, Quantel, France) -wavelength of 532 nm, 4 ns pulse duration, 10 Hz repetition rate -, were directed towards the sample. The beam was shaped by the combination of a lens (f’=75 mm, LA1608, Thorlabs, USA) and an engineered diffuser (EDC-10, RPC Photonics, USA) so that the laser illumination was homogeneous on the tube in the focal region of the transducer. The photoacoustic signals were detected with a 20 MHz spherically focused transducer (12.7 mm focal distance, 6.3 mm diameter, P120-2-R2.00, Olympus, Japan), amplified (DPR500, remote pulser RP-L2, JSR Ultrasonics, USA) and digitized with an oscilloscope (DLM 2024, Yokogawa).  A high frequency transducer was chosen to favor the observation of photoacoustic nonlinearities, as suggested by the theoretical predictions (sec. \ref{subsec:GenerationGNS}). 

Polycarbonate tubes with a 100 $\mu$m inner diameter (200 $\mu$m outer diameter, CTPC100-200, Paradigm Optics, USA) were used to approximately match the frequency of the emitted photoacoustic waves to that of the transducer. The tubes were filled with aqueous solutions of either 40-nm diameter gold nanospheres (HD.GC40.OD10, BBI Solutions, UK) or an organic blue dye (colorant E133, acidifier E330, preservative E202, ScrapCooking, France). The dilutions for each type of absorbers were set to obtain similar optical densities of about 10 ($\pm$ 0.1) at 532nm. 532 nm as chosen as the excitation wavelength as it is close to the absorption peak of gold nanospheres with diameters on the order of a few tens of nanometers.

For both samples, the peak-to-peak amplitude of the photoacoustic signal was studied as a function of either the incident fluence or the equilibrium temperature. The tubes were immersed in a large water tank providing a slow varying equilibrium temperature $T_{eq}$ when the water temperature was different from the room temperature. Temperature was monitored using a thermocouple (type K, RS Components, UK) placed in the vicinity of the tube, and connected to a computer via a thermocouple data loger (TC-08, Pico Technology, UK). A magnetic stirrer was used to maintain a homogeneous temperature distribution in the water tank. The tube was fixed on a beaker to prevent motion induced by water convection. A hot air flow was used to prevent condensation to form and perturb the laser beam on the entrance wall of the water tank. The incident fluence was controlled using a laser-integrated variable attenuator. For each position of the attenuator, the fluence values on the sample were deduced from a prior calibration and direct measurements of the average laser output powers using a thermal power meter. From a sample to the other, only the position of the tube was adjusted so as to be in the focus of the transducer and within the area where the calibration for the fluence remained valid. To improve the signal-to-noise ratio, the signals were coherently averaged on the oscilloscope ($N_{av}=32$), before the peak-to-peak value computed on the scope was transferred to the computer. The system noise was estimated from measurements with no illumination on the samples. 

For two equilibrium temperature $T_{eq}=4\degree C$ and $T_{eq}=20\degree C$, the fluence was varied from 0 mJ/cm$^2$ up to approximately 7 mJ/cm$^2$. For $T_{eq}=4\degree C$, water cooled in a freezer was used, the temperature of the tank remained around $4\pm 0.5\degree C$ during the few minutes of the experiment over which the incident fluence was varied. The experiments at $T_{eq}=20\degree C$ were done at room temperature. To study the evolution of the photoacoustic signal as a function of $T_{eq}$ from $2\degree C$ to $20\degree C$, the incident fluence was fixed at 5 mJ/cm$^2$, and the water tank initially filled with water from a freezer (at $T_{eq}< 2\degree C$). As the temperature slowly raised towards room temperature, both the water temperature and the photoacoustic signals were simultaneously measured and stored in the computer every 30 seconds. The same protocol was used for the two types of absorbing solutions (dye and gold nanospheres). For both samples, each type of experiment was repeated at least 3  times, and turned out to be reproducible.

\section{Results and Discussion}
\label{sec:ResultsDiscussion}

\begin{figure}[!h]
\includegraphics{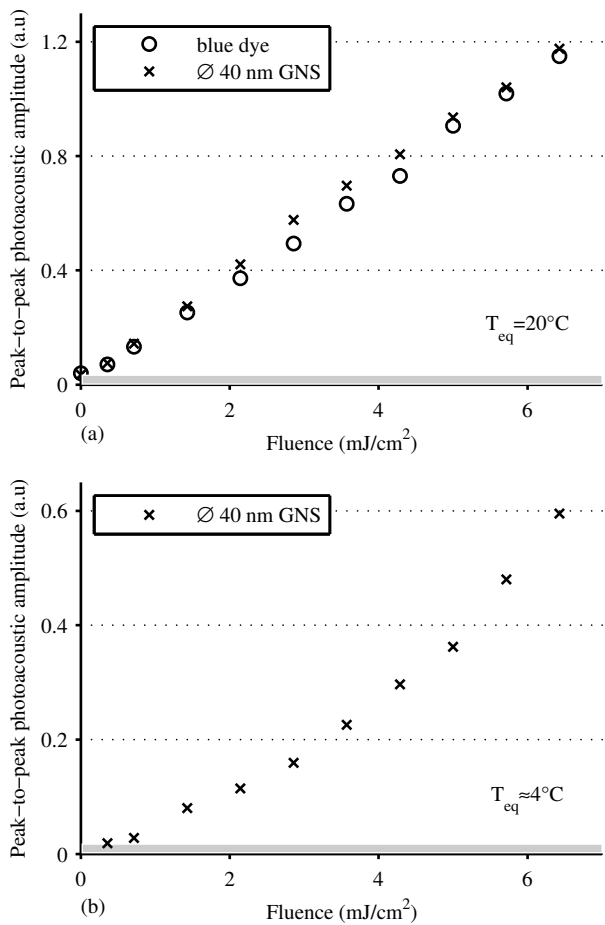}
\caption{Experimental values of the peak-to-peak photoacoustic amplitude as a function of the incident fluence. (a) Results obtained for $T_{eq}=20\degree C$: both absorbers behave linearly. (b) Results obtained for $T_{eq}\sim4\degree C$: no signal is detected from the dye sample, while a nonlinear signal is detected from the sample with gold nanospheres.}
\label{fig:fig4}
\end{figure}
Figure~\ref{fig:fig4}(a) shows typical results obtained for $T_{eq}=20\degree C$. No nonlinearity is observed for both types of absorbers as the peak-to-peak amplitude of the photoacoustic increased linearly with the fluence. However, when the temperature was approximately $T_{eq}=4\degree C$, no photoacoustic signal could be detected from the dye solution, while a photoacoustic signal well above the noise floor could be observed from the solution of gold nanospheres. Moreover, the photoacoustic amplitude from the gold nanospheres shows a clear nonlinear dependence with the incident fluence. At $T_{eq}=20\degree C$, the fact that a linear behavior was observed as a function of fluence, as opposed to what is predicted from Fig.~\ref{fig:fig2}(a), is likely due to the fact that the detection bandwidth centered around 20 MHz is not high enough to observe a significant nonlinear contribution. Fig.~\ref{fig:fig2}(a) predicts the peak-to-peak evolution of the unfiltered signal, whereas in the experiment the photoacoustic signals are filtered by both the detection bandwidth and the spatial extension of the excited absorbing solution. On the other hand, the fact that a photoacoustic signal of significant amplitude is observed with gold nanosphere at $T_{eq}\sim4\degree C$, while under the same condition none is observed from the dye solution, can be explained by the dynamic change of $\beta(T)$ around the nanosphere caused by the significant temperature rise. In this case, although the nonlinear generation is predominantly centered at much higher frequency than the detection frequency, it is the vanishing linear contribution at $T_{eq}\sim4\degree C$ that makes the nonlinear contribution detectable. For $T_{eq}=20\degree C$, the nonlinear contribution is probably too small compared to the linear one .

\begin{figure}[!h]
\includegraphics{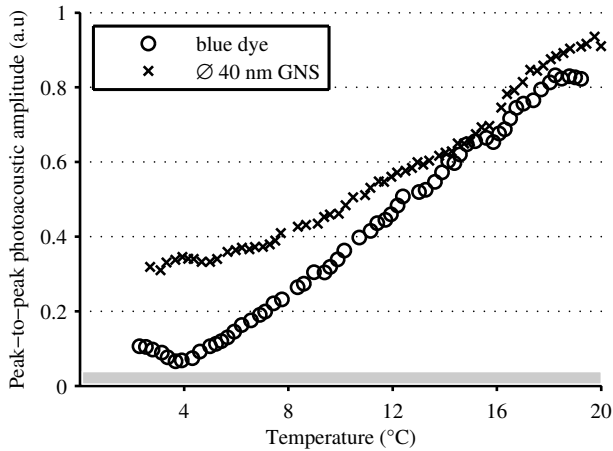}
\caption{Experimental peak-to-peak photoacoustic amplitude measured as a function of the equilibrium temperature $T_{eq}$, for an incident fluence of 5 mJ/cm$^2$.}
\label{fig:fig5}
\end{figure}

When the photoacoustic signal was measured as a function of the equilibrium temperature, as illustrated on Fig.\ref{fig:fig5}, for a fluence of 5 mJ/cm$^2$ (at which a nonlinear contribution was seen at $T_{eq}\sim 4\degree C$ from the gold nanospheres), the experimental results agreed with the predictions from Fig.\ref{fig:fig2}(c): the temperature dependence of the photoacoustic signal is significantly affected by the nanoscale local heating of the water surrounding the nanoparticles.

Several conclusions can be drawn from these experimental results. First, two absorbing media with the same optical density can be discriminated either from the dependence of photoacoustic amplitude with the incident fluence or from its dependence with the equilibrium temperature, if significant change of the coefficient of thermal expansion takes places during the laser illumination for one of them. Thermal based photoacoustic nonlinearity could be easily observed with gold nanospheres in water, with a detection frequency bandwidth centered around 20 MHz, at relatively low fluence provided that the temperature is close enough to $4\degree C$ so that the linear contribution is negligible.  The relevance of these results to biomedical applications requires further work but could be envisioned for imaging of biopsies. To date, only one experimental study reported the observation of presumably thermal based photoacoustic nonlinearity~\citep{Nam2012}. However, the required temperature rise was suggested to arise from the aggregation of nanoparticles within cells which led to some heat confinement, rather than from temperature rise around individual nanoparticules as considered in our work. It may be hypothesized that one reason for the lack of  studies reporting thermal-based photoacoustic nonlinearity in tissue with nanoparticles may be that most experiments used rather low frequency (in the MHz range), in combination with the limitation on fluence required by safety considerations (typically 20 mJ/cm$^2$). Further work needs to be done to assess whether it can be possible to detect nonlinear contribution in tissue at physiological temperature ($T_{eq}\sim 37\degree C$) by using a higher frequency ultrasound detection. Photoacoustic systems operating at frequencies up to 125 MHz have recently been reported~\citep{Zhang2012,Omar2013}.  For \textit{in vitro} application, our results suggest that if biological tissue behaves as water, high frequency photoacoustic  microscopy realized at a temperature close to $4 \degree C$ may allow discriminating nanoparticles from endogenous absorption, a feature that may be interesting for experimental studies of particle uptake in tissue for instance. Our study also suggests that the temperature dependence of photoacoustic amplitude, a feature that has been proposed as a mean to monitor temperature changes in tissue during thermal therapies~\citep{Larina2005,Shah2008}, may not only depend on the type of tissue, but if thermal nonlinearities are involved, may also depend on both the incident fluence and the type of absorbers used to photoacoustically probe the medium. Further works is also required to check the relevance of this to \textit{in vivo} conditions.

\section{Conclusions}

In this work, we demonstrated that thermal-based photoacoustic nonlinearities can be measured using a 20 MHz detection frequency with 40-nm gold nanospheres in water at relatively low fluence. The nonlinearity is observed either as a nonlinear dependence of the photoacoustic amplitude against the incident fluence, or as a fluence- and absorber- dependent relationship between the photoacoustic amplitude and the equilibrium temperature. As a consequence, two solutions of identical optical density were discriminated based on the effect of the nonlinearity. The experimental observations agree qualitatively with theoretical predictions based on the temperature-dependence of the coefficient of thermal expansion in water.

\section{Acknowledgments}
The authors thank Mickaël Mounaix for his assistance with the reproducibility experiments. Olivier Simandoux gratefully acknowledges funding from the Direction Générale de l’Armement (DGA). This work was funded by grants from the ANR (Golden Eye), the INCA (Gold Fever), and CNRS-PSL*.

\section{References}

\bibliographystyle{model1-num-names}

\end{document}